\documentclass{emulateapj}
\usepackage{float}
\usepackage{amsmath}
\usepackage{graphicx}

\begin{document}

\title{Compact E+A Galaxies as a Progenitor of Massive Compact Quiescent Galaxies at $0.2<\lowercase{z} < 0.8$}
\author{H. Jabran Zahid$^{1}$, Nicholas Baeza Hochmuth$^{1,2}$, Margaret J. Geller$^{1}$, Ivana Damjanov$^{3}$, Igor V. Chilingarian$^{1,4}$, Jubee Sohn$^{1}$, Fadia Salmi$^{5}$ and Ho Seong Hwang$^{5}$}
\affil{$^{1}$Smithsonian Astrophysical Observatory, Harvard-Smithsonian Center for Astrophysics$-$60 Garden St., Cambridge, MA 02138, USA}
\affil{$^{2}$Department of Physics \& Applied Physics, University of Massachusetts, Lowell$-$1 University Ave., Lowell, MA 01854, USA}
\affil{$^{3}$Harvard College Observatory, Harvard-Smithsonian Center for Astrophysics$-$60 Garden St., Cambridge, MA 02138, USA}
\affil{$^{4}$Sternberg Astronomical Institute, M.V.Lomonosov Moscow State University, 13 Universitetsky prospect, Moscow 119992 Russia}
\affil{$^{5}$School of Physics, Korea Institute for Advanced Study$-$85 Hoegiro, Dongdaemun-Gu, 02455 Seoul, Korea}

\begin{abstract}

We search the Sloan Digital Sky Survey and the Baryon Oscillation Sky Survey to identify $\sim5500$ massive compact quiescent galaxy candidates at $0.2<z<0.8$. We robustly classify a subsample of 438 E+A galaxies based on their spectral properties and make this catalog publicly available. We examine sizes, stellar population ages and kinematics of galaxies in the sample and show that the physical properties of compact E+A galaxies suggest that they are a progenitor of massive compact quiescent galaxies. Thus, two classes of objects$-$compact E+A and compact quiescent galaxies$-$may be linked by a common {formation scenario}. The typical stellar population age of compact E+A galaxies is $<1$ Gyr. The existence of compact E+A galaxies with young stellar populations at $0.2<z<0.8$ means that some compact quiescent galaxies first appear at intermediate redshifts. We derive a lower limit for the number density of compact E+A galaxies. Assuming passive evolution, we convert this number density into an appearance rate of new compact quiescent galaxies at $0.2<z<0.8$. The lower limit number density of compact quiescent galaxies which may appear at $z<0.8$ is comparable to the lower limit of the total number density of compact quiescent galaxies at these intermediate redshifts. Thus, a substantial fraction of the $z<0.8$ massive compact quiescent galaxy population may descend from compact E+A galaxies at intermediate redshifts.

\end{abstract}
\keywords{galaxies: formation --- galaxies: evolution --- galaxies: starburst --- galaxies: fundamental parameters}

\section{Introduction}

The formation and growth of galaxies is the subject of intense research. The average size of massive galaxies which have ceased star formation (i.e., quiescent galaxies) is substantially smaller at $z\sim2$ than similarly massive systems in the nearby universe \citep[e.g.,][]{Daddi2005, vanderwel2014}. The physical origin for the increase in the average size of quiescent galaxies remains uncertain and various mechanisms have been proposed. Individual galaxies may grow via mergers \citep[e.g.,][]{White2007, Bezanson2009, Naab2009, Newman2012}, {by the acquisition of disks at later times \citep{Dullo2013, Graham2015}} or galaxies added to the quiescent population at later times may be larger \citep[i.e., progenitor bias;][]{vanDokkum2001, vanderwel2009, Carollo2013}. The growth of individual compact quiescent galaxies and progenitor bias are not mutually exclusive explanations for the average increase in size of the quiescent galaxy population at late times \citep[e.g.,][]{Khochfar2006}.

The abundance of massive compact quiescent galaxies may help to discriminate among various growth scenarios. The number density of compact galaxies in the nearby universe measured from the Sloan Digital Sky Survey (SDSS) appears to be a few orders of magnitude smaller than at $z\sim2$ \citep{Trujillo2009, Taylor2010}. However, several recent studies argue that the abundance of compact galaxies is nearly constant when dense regions in the local universe \citep{Valentinuzzi2010, Poggianti2013} and/or intermediate redshifts are probed \citep{Carollo2013, Damjanov2014, Damjanov2015a}. There is some {possible} tension between the increasing average size of quiescent galaxies with time and the constant abundance of compact galaxies at $z<1$. The appearance of new massive compact quiescent galaxies at $z<1$ would mitigate this tension. 

E+A galaxies have spectral properties of a young stellar population superposed on an older population \citep{Dressler1983}. These galaxies are defined by strong Balmer absorption lines and the absence of emission lines; properties indicative of a starburst in the last $\lesssim1$ Gyr and a cessation of star formation shortly thereafter \citep[e.g.,][and many others]{Dressler1983, Couch1987, Fabricant1991, Dressler1992, Tran2003, Tran2004, Goto2005}. Thus, E+A galaxies are a class of objects which can be spectroscopically identified as newly formed quiescent galaxies, i.e. galaxies which have only recently shut down star formation.

The E+A phase is short-lived. Whether a galaxy which has recently shut down star formation experiences an E+A phase depends on the star formation history and transformation scenario \citep{Falkenberg2009}. The spectra of many high redshift compact quiescent galaxies show the standard spectral features of E+As \citep[e.g., see Figure 1 in][]{vanDokkum2008b, Kriek2006, Bezanson2013b}. Similar spectral features are also observed in some intermediate and low redshift compact quiescent galaxies \citep{Trujillo2009, Damjanov2014}. Thus, the E+A phase may be a common evolutionary stage for compact quiescent galaxies. 

Compact E+A galaxies are a subset of the broader class of compact quiescent galaxies. Whether compact E+A galaxies (which are quiescent by definition) are included in samples of compact quiescent galaxies depends on the definition of quiescence. Galaxies are often classified as quiescent based on photometry rather than spectroscopy. Colors of galaxies are a global property of the stellar population which evolve more slowly than emission line properties which depend on massive, short-lived stars. Thus, samples of compact quiescent galaxies identified photometrically may or may not include compact E+A galaxies. Compact E+A galaxies do however appear in samples of compact quiescent galaxies identified spectroscopically \citep[e.g.,][]{Damjanov2014}.

We seek to establish and quantify the appearance of new massive compact quiescent galaxies at intermediate redshifts. Within the massive compact quiescent galaxy population, we search for the subset of compact E+A galaxies. We classify E+A galaxies in a sample of massive compact quiescent galaxies with SDSS and Baryon Oscillation Sky Survey (BOSS) spectroscopy at $0.2<z<0.8$. Compact E+A galaxies at $z<0.8$ {would} suggest that new massive compact quiescent galaxies do appear at intermediate redshifts. Quantifying the abundance of compact E+A galaxies is critical for understanding how new compact quiescent galaxies contribute to the preexisting compact quiescent galaxy population at intermediate redshifts.

The data and methods are described in Section 2. The physical properties of galaxies are {presented} in Section 3 and we derive an estimate for the appearance rate of new compact galaxies in Section 4. We discuss the results in Section 5 and conclude in Section 6. Throughout we assume the standard cosmology $(H_{0}, \Omega_{m}, \Omega_{\Lambda})$ = (70 km s$^{-1}$ Mpc$^{-1}$, 0.3, 0.7).

\section{Data and Methods}

We select objects from the SDSS DR12\footnote{http://www.sdss.org/dr12/} \citep{Alam2015} which includes resolved sources from the Legacy survey of $\sim900,000$ galaxies with $r<17.8$ \citep{York2000} and the Baryon Oscillation Spectroscopic Survey (BOSS) of $\sim1.5$ million luminous red galaxies with $i < 19.9$ \citep{Eisenstein2011}. The surveys also include $\sim350,000$ unresolved sources targeted as quasars in the SDSS and BOSS with $i<20.2$ and $g<22$, respectively \citep{Richards2002, Ross2012}. {A small fraction of the unresolved sources in our sample are from the Sloan Extension for Galactic Understanding and Exploration (SEGUE) stellar survey \citep{Yanny2009}. }

The nominal spectral range of the SDSS and BOSS observations are $3800 - 9200 \mathrm{\AA}$ and $3600-10000\mathrm{\AA}$, respectively, at a spectral resolution of $R\sim2000$. For the resolved sources, we adopt the Petrosian $i$-band half-light radius measured from the standard SDSS photometric reduction \citep{Stoughton2002}. For the unresolved sources, we use the PSF magnitudes from the photometric pipeline \citep{Stoughton2002}. We use the stellar masses, emission line properties and velocity dispersions calculated by the Portsmouth Group\footnote{http://www.sdss.org/dr12/spectro/galaxy\textunderscore portsmouth/}. Stellar masses are derived by fitting the $ugriz$ magnitudes assuming a \citet{Kroupa2001} initial mass function (IMF) and the stellar population models of \citet{Maraston2005}. Details of the velocity dispersion and emission line measurements are in \citet{Thomas2013}. 

We use the surveys to identify massive compact quiescent galaxies and to search for a subsample of post-starburst galaxies at $0.2<z<0.8$. We derive two parent samples from the SDSS DR12. We select the first sample from resolved objects in the SDSS and BOSS that are compact on the basis of SDSS imaging. We select the second sample from unresolved objects targeted as stars or quasars. Both samples are limited to $0.2<z<0.8$. The redshift selection yields a sample of bright, high surface brightness candidates. These objects tend to be massive compact galaxies \citep{Zahid2015}. Quiescent, non-star-forming galaxies are selected throughout by requiring that the equivalent width of [OII]$\lambda3726$ + [OII]$\lambda3728$ (EW [OII]) , EW H$\beta$ and EW H$\alpha$ is $< 2.5\mathrm{\AA}$.

\subsection{The Resolved Sample}

We select the resolved sample from the main galaxy sample of the SDSS and the luminous red galaxy sample of BOSS. {We refer to this sample as the ``resolved parent sample."} We select compact objects based on a widely used classification of compactness \citep[B13 hereafter]{Barro2013}:
\begin{equation}
\Sigma_{1.5} = \mathrm{log}\left(M / R_e^{1.5} \right) > 10.3 \, \, M_\odot \, \, \mathrm{kpc}^{-1.5}.
\label{eq:b13}
\end{equation}
Here, $M$ is the stellar mass and $R_e$ is the half-light radius. The B13 classification is based on stellar masses calculated assuming the \citet{Chabrier2003} IMF which are $\sim0.05$ dex smaller than stellar masses used here (calculated based on the Kroupa IMF). However, this difference is negligible because the classification of compact galaxies based on SDSS sizes is very conservative. 

The SDSS Petrosian half-light radius is not corrected for seeing, typically $\gtrsim1''.2$ \citep{Stoughton2002}. Thus, the SDSS Petrosian sizes of galaxies$-$which are generally near the resolution limit$-$are significantly overestimated; consequently $\Sigma_{1.5}$ is significantly underestimated \citep[e.g.,][]{Shen2003}.  

{We show below that the $\Sigma_{1.5}$ selection criteria we apply robustly select massive compact quiescent galaxies in the SDSS. Using the selection criteria we derive a {resolved} parent sample of 3809 massive compact quiescent galaxies (1526 and 2283 galaxies in the parent sample are part of the SDSS Legacy and BOSS luminous red galaxies samples, respectively). Figure \ref{fig:hist} shows the properties of the resolved parent sample.}

\begin{figure}
\begin{center}
\includegraphics[width=\columnwidth]{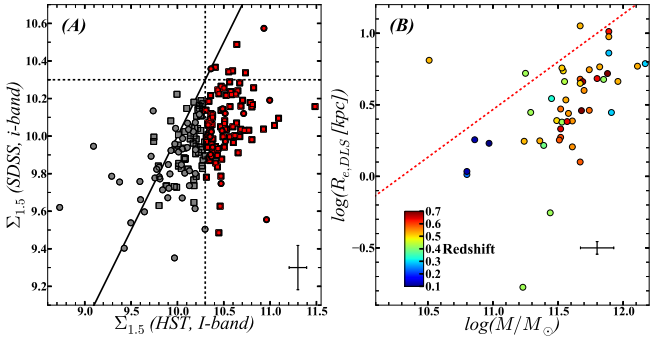}
\end{center}
\caption{(A) Compact classification of the same 202 galaxies based on SDSS and HST sizes \citep{Sargent2007, Griffith2012}. The dotted lines show the compact galaxy classification of B13 and the solid line is the one-to-one relation. The typical error is shown in the lower right. Due to the lack of uniform spectroscopy for this sample, quiescent galaxies are selected based on rest-frame colors. (B) The size versus mass diagram for {47} galaxies classified as compact based on SDSS photometry but with sizes remeasured using higher quality imaging from the DLS (see text for detail). The red dotted line is the B13 compact galaxy classification. The typical error bars are shown in the lower right. For this comparison, quiescent galaxies are selected based on $D_{n}4000$ ($>1.5$).}
\label{fig:size}
\end{figure}

\begin{figure*}
\begin{center}
\includegraphics[width=2\columnwidth]{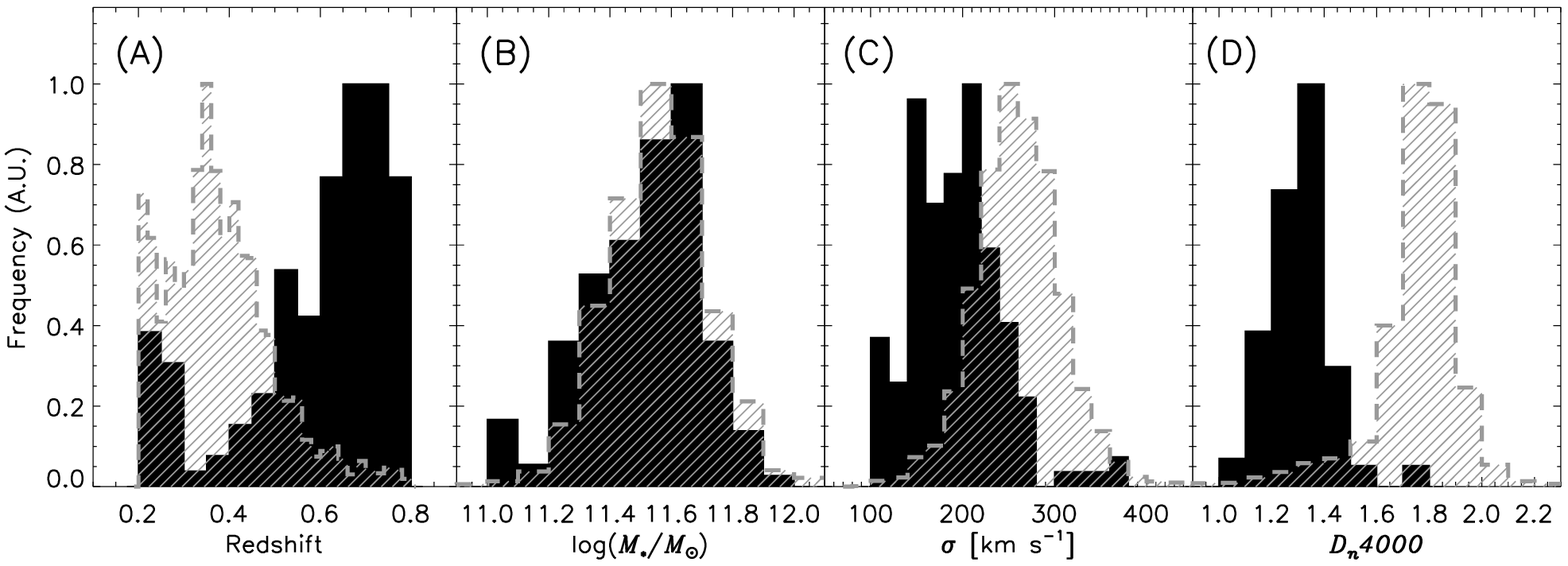}
\end{center}
\caption{(A) Redshift, (B) stellar mass (C) velocity dispersion and (D) $D_{n}4000$ distribution of the resolved parent (gray dashed, hashed) and E+A samples (solid black). The peak value of the histograms are normalized to unity. {We abbreviate arbitrary units as A.U. throughout.}}
\label{fig:hist}
\end{figure*}

\subsection{Testing the Compact Galaxy Classification}

{We test the robustness of the classification of compact galaxies by comparing sizes measured for the same objects using both SDSS and HST imaging. We cross match SDSS galaxies with catalogs of HST sizes using a 0''\!\!.5 matching radius. These galaxies are not part of the sample we analyze in this study; this sample of galaxies with SDSS and HST imaging is derived solely to test the robustness of the compact galaxy classification which is based on SDSS imaging alone. We identify 143 and 59 galaxies in the \citet{Sargent2007} and \citet{Griffith2012} catalogs, respectively. For this sample, 100 galaxies have spectroscopic redshifts from various sources and 102 have only photometric redshifts \citep{Ilbert2009}. We refer to this cross-matched sample as the ``overlap" sample.}

The HST size estimates {for the overlap sample} are based on imaging from the Advanced Camera for Surveys. Each galaxy size is based on a Sersic model profile \citep{Sersic1968, Graham2005}:
\begin{equation}
\Sigma(R) = \Sigma_e e ^{ \left( -k[(R/R_e)^{1/n}-1] \right) }.
\label{eq:sersic}
\end{equation}
The free parameters of the model are the half-light radius, $R_e$ and the Sersic index, $n$. $\Sigma_e$ is the surface brightness at $R_e$ and $k$ is a constant chosen such that half the light is contained within $R_e$. To account for the point spread function (PSF), each two dimensional model is first convolved with the PSF derived from observations of bright stars and then fit to the observations.

One hundred and three of the 202 objects in the {overlap sample} are compact on the basis of HST sizes. Figure \ref{fig:size}A shows a comparison of $\Sigma_{1.5}$ calculated from SDSS and HST. Most galaxies fall below the one-to-one relation because SDSS sizes are not seeing corrected and therefore typically are an overestimate the true size of small galaxies. Many bona fide compact galaxies (red points in Figure \ref{fig:size}A) do not appear compact based on SDSS sizes. Figure \ref{fig:size}A also shows that every galaxy that is classified as compact based on SDSS imaging is also compact based on HST imaging.

In Figure \ref{fig:size}B we test the compact classification further. Based on SDSS imaging, we identify compact galaxies with $\Sigma_{1.5} > 10.3$ in the four square degree F2 field of the Deep Lensing Survey \citep[DLS,][]{Wittman2002}. We remeasure sizes for these galaxies with deeper, higher resolution DLS photometry using GALFIT \citep{Peng2002} and the Sersic model described by Equation \ref{eq:sersic}. The images are taken when seeing is $\leq0''.9$ \citep{Wittman2002} and the sizes are corrected for seeing by convolving models with the PSF determined from observations of stars in the DLS images. A significant majority of objects (94\%) classified as compact based on SDSS sizes are also classified as compact based on DLS sizes. 

Figures \ref{fig:size}A and \ref{fig:size}B both show that SDSS sizes are overestimated; thus the compact galaxy classification we use for the resolved sample is quite conservative.

\subsection{The Point Source Sample}

\begin{figure*}
\begin{center}
\includegraphics[width=2\columnwidth]{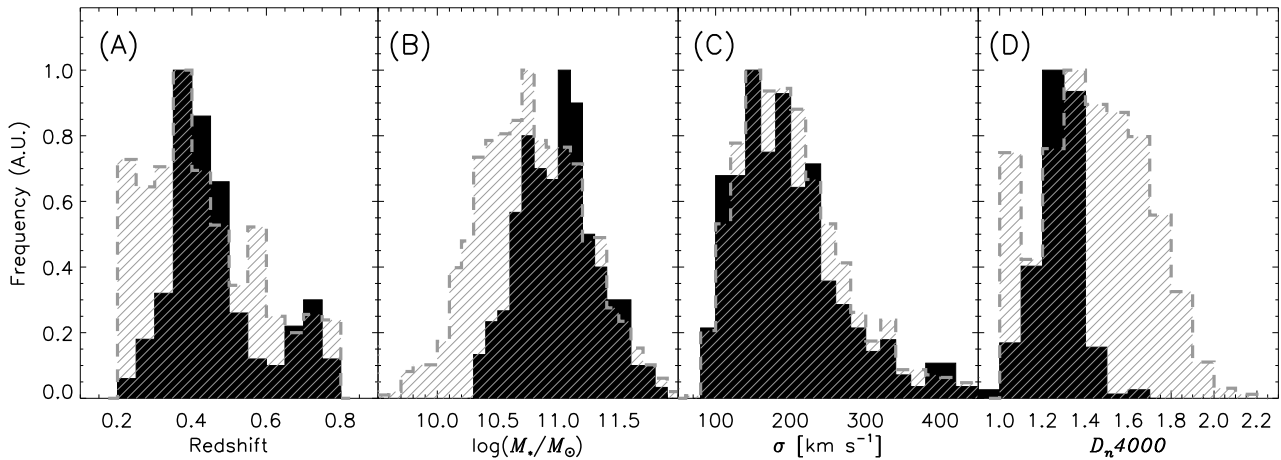}
\end{center}
\caption{(A) Redshift, (B) stellar mass (C) velocity dispersion and (D) $D_{n}4000$ distribution of the point source parent (gray dashed, hashed) and E+A samples (solid black). The peak value of the histograms are normalized to unity.}
\label{fig:hist_point}
\end{figure*}

We select a second parent sample from galaxies classified as point sources in SDSS imaging. {We refer to this sample as the ``point source parent sample."} Most of the objects were initially targeted as quasars \citep{Richards2002, Ross2012}. We first select all objects with photometric flag ``TYPE" equal to ``STAR". This selection limits the sample to unresolved objects. We then select objects with the spectroscopic flag ``CLASS" equal to ``GALAXY" removing objects identified as quasars and stars. The point source {parent} sample consists of 1714 galaxies: 1001, 589 and 124 of the galaxies are from the BOSS, SDSS and SEGUE surveys, respectively.

\citet{Damjanov2014} examine point sources identified as galaxies in BOSS to estimate the number density of compact galaxies at intermediate redshifts. Fourteen of the point sources in their sample are found in the Canada-France-Hawaii Telescope (CFHT) MegaCam database. The CFHT MegaCam images are deeper and observed at higher resolution (seeing $<0."9$). \citet{Damjanov2014} fit the Sersic profile described by Equation \ref{eq:sersic} using GALFIT. The final fits are corrected for the PSF derived from point sources near each target galaxy. All 14 of the galaxies that are point sources in the SDSS imaging are compact galaxies based on the CFHT MegaCam imaging. \citet{Damjanov2014} conclude that $>82\%$ of the galaxies in their point source sample are compact. We select our sample in a similar manner. Thus, the vast majority of these unresolved galaxies are compact.

\begin{deluxetable*}{lcccc}
\tablecaption{Sample Source}
\tablehead{\colhead{Sample} &\colhead{BOSS} &\colhead{SDSS} &\colhead{SEGUE } & \colhead{TOTAL} }
\startdata
Resolved Parent  Sample   &   2283 & 1526 & 0 & 3809 \\
Resolved E+A  Sub-Sample & 109 & 39 & 0 & 148 \\
Point Source Parent  Sample  & 1001 & 589  & 124 & 1714\\
Point Source E+A Sub-Sample  & 144 & 144 & 2 & 290 \\
Total Sample  & 3284 & 2115 & 124 & 5523 \\
\enddata
\label{tab:sample}
\end{deluxetable*}

\subsection{E+A Galaxy Identification}

We adopt a standard definition of E+A galaxies: [OII] EW $< 2.5 \mathrm{\AA}$ and H$\delta$ EW $< -5 \mathrm{\AA}$\footnote{Negative EWs indicate absorption lines.}. We follow the procedure outlined in \citet{Goto2003} to measure H$\delta$ EWs. We use the H$\delta$ (wide) wavelength windows. The blue and red continuum are $4030 - 4082 \mathrm{\AA}$ and $4122-4170 \mathrm{\AA}$, respectively. The wavelength range for the line is $4082 - 4122 \mathrm{\AA}$. A linear chi-square fit to the flux in the two continuum bands determines the continuum. The residuals from the continuum fit are calculated, 3$\sigma$ outliers are removed and the continuum is refit. The flux in the line wavelength window is normalized to the best-fit continuum. The EW of the line is determined by summing the pixels in the line wavelength window. The EW error is determined by standard error propagation of the flux uncertainties.

We measure the H$\delta$ EW for all galaxies. For a robust identification, we limit the E+A sample to galaxies with H$\delta$ EW measured with a signal-to-noise (S/N) ratio $>3$. A small fraction of objects are spurious due to strong sky line contamination and/or unidentified emission lines. We identify these objects by visual examination.

The final sample consists of 438 E+A galaxies: 148 and 290 are from the resolved and point source parent samples, respectively. Of the 148 galaxies in the resolved sample, 109 and 39 are part of BOSS and SDSS, respectively. Of the 290 galaxies in the point source sample, 144, 144 and 2 are in the BOSS, SDSS and SEGUE samples, respectively. Table 1 summarizes the various samples and the source surveys and Figures \ref{fig:hist} and \ref{fig:hist_point} show their properties.

Our selection and identification of compact E+A galaxies is complementary to previous catalogs of E+A galaxies from the SDSS. There is no overlap with the catalog of T. Goto\footnote{http://www.phys.nthu.edu.tw/$\sim$tomo/cv/index.html} which is based on the DR7 and is selected with S/N$>10$ for absorption lines. The E+A properties are in Table 3.

\subsection{Stellar Population Ages}

The $D_n4000$ index is the flux ratio between two spectral windows adjacent to the 4000$\mathrm{\AA}$ break \citep{Balogh1999}. We measure the $D_n4000$ index directly from the SDSS spectra. The index is related to the age of the stellar population \citep{Kauffmann2003a} and we use it as a directly measured proxy of the relative age. 

We estimate ages of galaxies by fitting their spectra against a grid of simple stellar population (SSP) models computed with the {\sc pegase.hr} code \citep{LeBorgne2004} using the {\sc nbursts} full spectrum fitting technique \citep{Chilingarian2007a, Chilingarian2007b}. We limit the spectral fit to the rest-frame wavelength range of $3700-6800\mathrm{\AA}$. The ages we measure represent the luminosity weighted SSP-equivalent value (i.e., a model with a single instantaneous starburst).

\begin{figure*}
\begin{center}
\includegraphics[width=1.5\columnwidth]{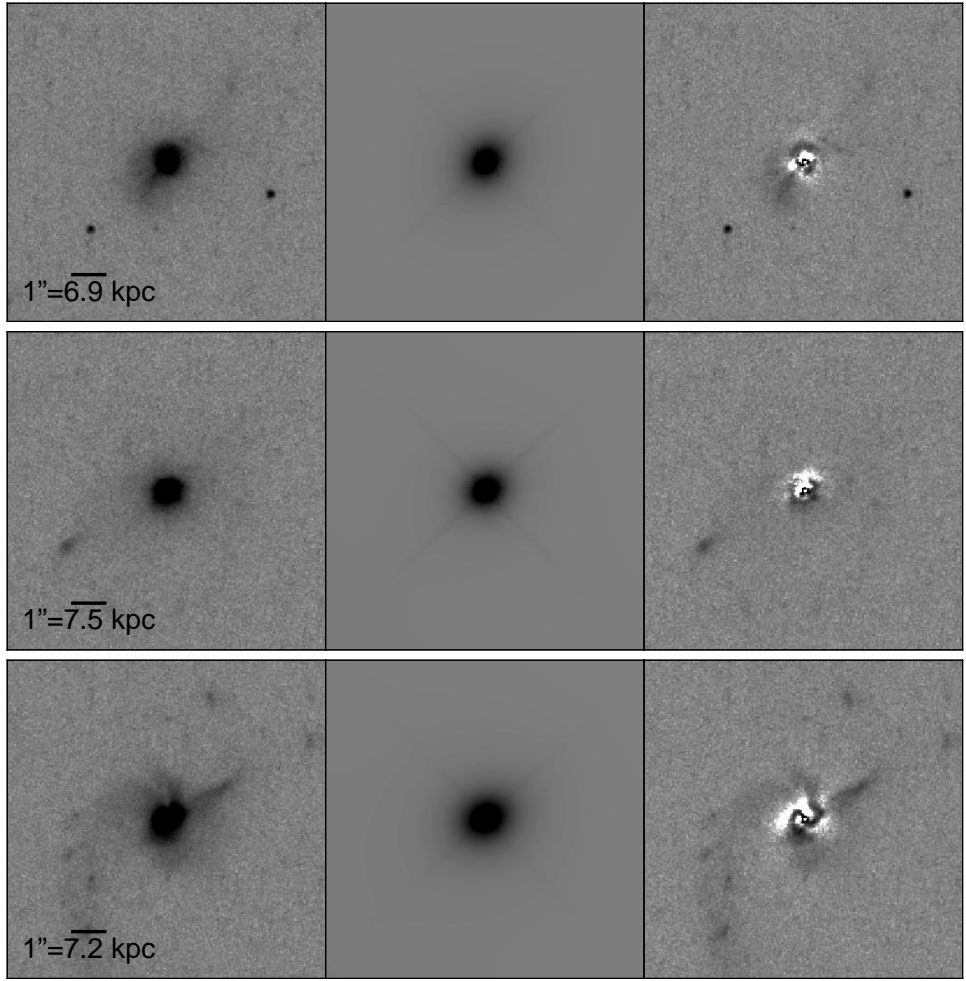}
\end{center}
\caption{HST images for three E+A galaxies. The first column shows a thumbnail ACS image of the galaxy. The second column shows the best-fit two dimensional Sersic model (Equation \ref{eq:sersic}). The third column shows the residuals. A fourth galaxy imaged by HST is unresolved and thus not shown here.}
\label{fig:hst}
\end{figure*}

Figures \ref{fig:hist} and \ref{fig:hist_point} show the measured properties of the E+A galaxies. For the resolved sample, the distribution of the E+A galaxy properties differs significantly from the parent sample highlighting the selection bias present in that sample. For the point source sample, the distribution of the E+A galaxy and parent sample properties are more similar. Figure \ref{fig:hist_point}B shows that the stellar masses of E+A galaxies are shifted towards more massive objects. This shift is largely a consequence of the higher S/N spectroscopy required to identify E+A galaxies that biases the distribution towards brighter, more massive objects. {The difference in the stellar mass distributions of the resolved and point source parent samples (Figures \ref{fig:hist}B and \ref{fig:hist_point}B) is due to the different magnitude limits of the surveys used to select these samples.}

\section{Properties of E+A Galaxies}

The compact E+A galaxies at $0.2<z<0.8$ have recently ceased star formation. Here, we examine the physical properties of these galaxies to explore the possibility that they are the progenitors of compact quiescent galaxies at intermediate redshifts. In Section 3.1 we examine the sizes of compact E+A galaxies measured from HST imaging and in Section 3.2 we compare the stellar population ages of compact E+A galaxies with the parent samples (resolved + point source) of compact quiescent galaxies. 

We examine the velocity dispersions and derive number densities in Sections 3.3 and 3.4, respectively. In these two sections, we restrict our analysis to the BOSS point source sample. The BOSS point source sample is the deepest, uniformly sampled survey covering a large fraction of the sky. The depth of the survey ensures that the sample contains a large number of sources across the redshift range we analyze; the large sky coverage mitigates the effects of cosmic variance. For the analysis in Sections 3.3 and 3.4, we do not use the other samples in this study because they are not uniformly sampled at the depth of BOSS.

\subsection{HST Sizes}

\begin{figure}
\begin{center}
\includegraphics[width=\columnwidth]{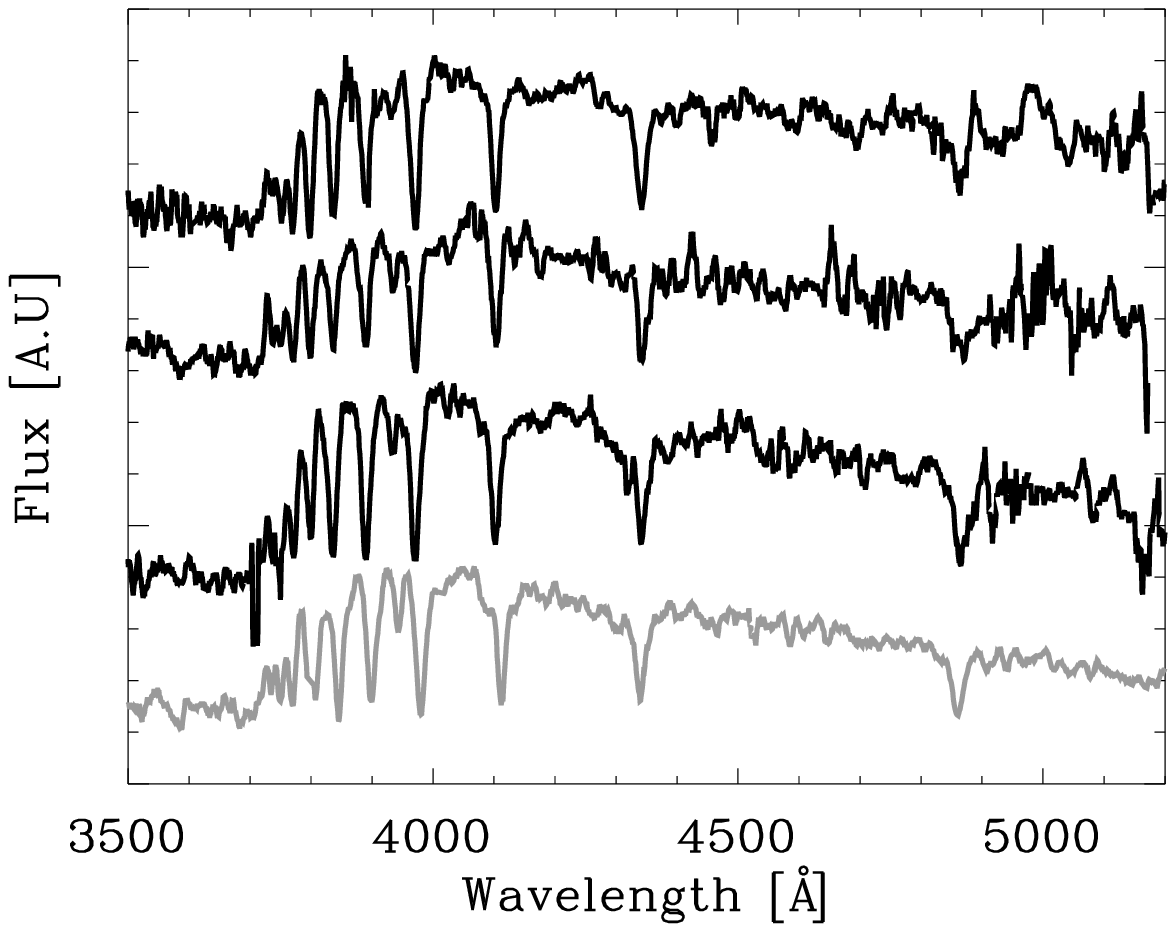}
\end{center}
\caption{Rest-frame spectra of the four galaxies observed with HST. The top three spectra correspond to the galaxies shown in Figure \ref{fig:hst} plotted in the same order from top to bottom. The gray spectrum is the E+A galaxy which is unresolved in HST WFC3 imaging. {The spectra are presented in the same order from top to bottom as the object properties given in Table \ref{tab:hst}.}}
\label{fig:spectra}
\end{figure}

\begin{deluxetable*}{cccccccl}
\tablecaption{HST Imaged Sample Properties}
\tablehead{\colhead{RA} &\colhead{Dec} &\colhead{Redshift} &\colhead{log$(M_\ast/M_\odot)$ } & \colhead{$R_e$ [kpc]}  & \colhead{$\Sigma_{1.5} [M_\odot \, \mathrm{kpc}^{1.5}]$ }  & \colhead{Age [Myr]} & \colhead{Sample}}
\startdata
192.02979    &   6.0199444 & 0.63229 & 11.58 & 0.99 & 11.59 & 960 & Point Source\\
222.62021     &  46.360472 & 0.78184 & 11.86 & 0.64 & 12.15 & 20  & Point Source\\
248.92379    &   47.156778 & 0.69931 & 11.84 & 1.59 & 11.54 & 270 & Resolved\\
198.74036    &   53.574875 &    0.39277 & 11.36  & $<0.70$ & $>11.59$ & 680 & Point Source\\
\enddata
\label{tab:hst}
\end{deluxetable*}

To establish that E+A galaxies are indeed compact, we cross-match our E+A catalog with the HST archive. Four galaxies in our sample are imaged with HST. Figure \ref{fig:hst} shows WFC3/UVIS F814W images of three of the four galaxies \citep[HST GO 12272; PI: C. Tremonti;][]{Diamond-Stanic2012}. The fourth galaxy is observed with the WFC3/IR F160W but is unresolved even in HST imaging (HST Program ID 12613, PI: K. Jahnke). For each of the resolved galaxies we fit the profile with GALFIT using the Sersic profile model given by Equation \ref{eq:sersic}. For the unresolved source we adopt the half-width half maximum of the WFC3 F160W PSF ($\sim0.''13$) as an upper limit on the galaxy size. The properties of the four objects are in Table 2.  

The $\Sigma_{1.5}$ we measure firmly establishes the four galaxies with HST imaging as compact. Figure \ref{fig:spectra} shows the spectra for these four galaxies. The spectra exhibit the defining characteristics of E+A galaxies; strong Balmer absorption and no measurable emission lines. While the cross-matched sample is small, all four galaxies we identify as compact E+As are indeed very compact based on the $\Sigma_{1.5}$ definition. The result that all four galaxies imaged with HST are compact is consistent with \citet{Damjanov2014} who apply a similar selection to identify compact galaxies in the BOSS. 



\subsection{Stellar Population Ages}

\begin{figure}
\begin{center}
\includegraphics[width=\columnwidth]{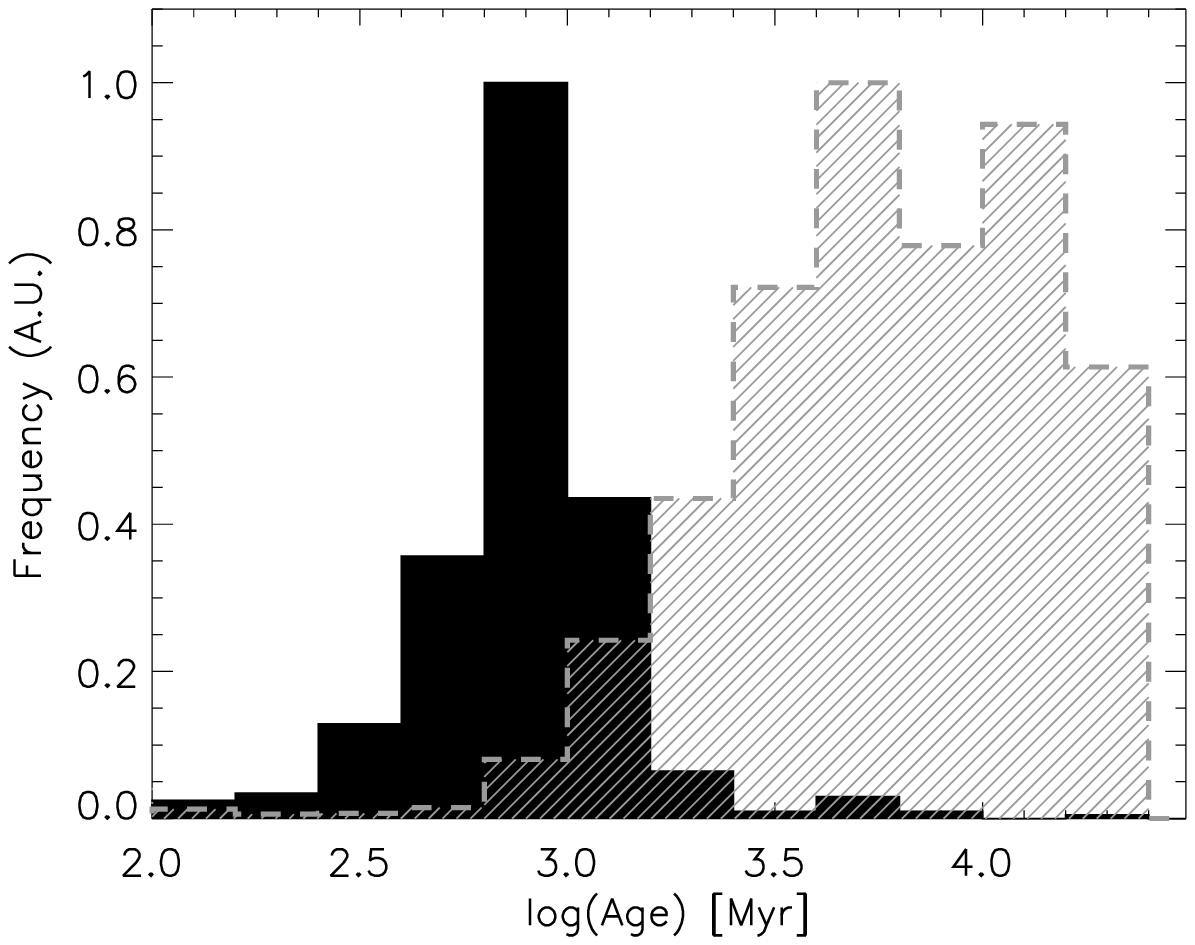}
\end{center}
\caption{{Single burst, simple stellar population equivalent age distribution determined from fitting models to the spectra of galaxies in the total sample. The solid black histogram is the age distribution of the E+A galaxies in the total sample (see text for sample definition). The gray, hashed histogram is the age distribution of galaxies in the complementary (excluding E+A galaxies) total sample.}}
\label{fig:age}
\end{figure}

\begin{figure}
\begin{center}
\includegraphics[width=\columnwidth]{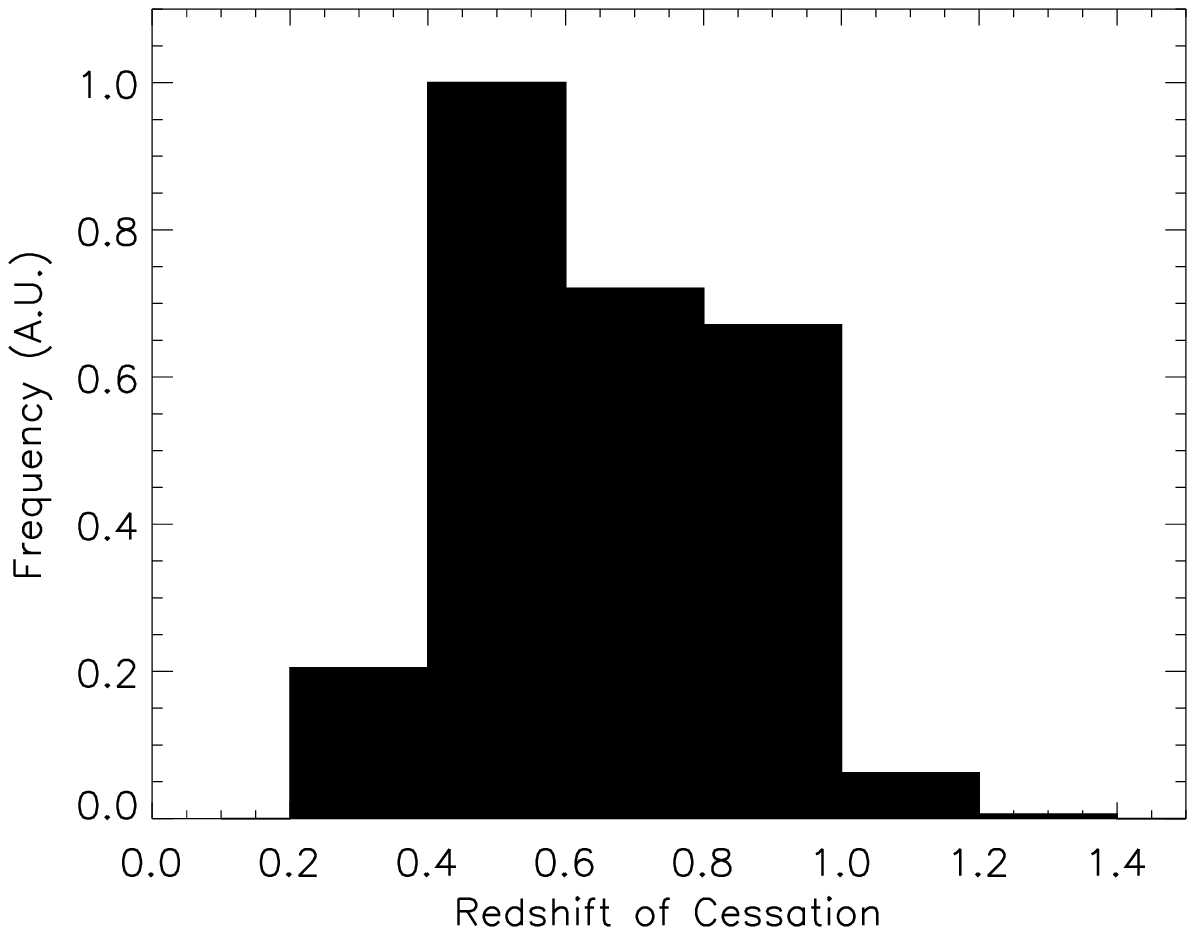}
\end{center}
\caption{``Redshift of cessation" for E+A galaxies in our combined sample.}
\label{fig:zform}
\end{figure}

Here we examine the stellar population ages of E+A and massive compact quiescent galaxies to assess when these two populations of galaxies shut down star formation and became quiescent.

Figures \ref{fig:hist}D and \ref{fig:hist_point}D show the $D_{n}4000$ index distribution of the E+A and parent samples of quiescent galaxies. The $D_{n}4000$ indices of the E+A galaxies are $\lesssim1.5$. The $D_{n}4000$ distribution suggests that the stellar populations of E+A galaxies in our sample are typically younger than the parent population, consistent with the recent cessation of star formation in E+A galaxies.

Figure \ref{fig:age} shows the stellar population age distribution for galaxies in {the resolved and point source parent samples combined (which we refer to as the ``total" sample)}. Although the typical timescale for the starburst associated with E+A galaxies is $<1$ Gyr, the SSP-equivalent age is a luminosity weighted average over the whole stellar population; thus, older stellar populations present at the time of the starburst are averaged into the age estimate. Figure \ref{fig:age} shows that E+A galaxies have SSP-equivalent ages that are $<3$ Gyr with the peak of the distribution around $\sim800$ Myr. The parent sample tends to be older with ages distributed around 5-10 Gyr. The comparison of ages supports our conclusion based on the $D_{n}4000$ distribution that E+A galaxies represent the young tail of the quiescent galaxy age distribution.

We estimate the redshift of the E+A starburst by subtracting the SSP-equivalent age\footnote{For E+A galaxies we treat the measured luminosity weighted SSP-equivalent age as an upper limit estimate of the age of the starburst; older stellar populations contribute to the luminosity and thus to the age we derive for the stellar population.} from the age of the universe at the observed redshift of each galaxy. Because the timescale of the starburst is significantly shorter than the stellar population ages we measure \citep{Mihos1996}, the starburst redshift is roughly coincident with the redshift where star-formation shut down. We refer to this redshift as the ``redshift of cessation" and plot the distribution in Figure \ref{fig:zform}. The massive compact E+A galaxies are significantly younger than the parent population and the vast majority ceased forming stars at $z<1$.

\subsection{Velocity Dispersions}

\begin{figure}
\begin{center}
\includegraphics[width=\columnwidth]{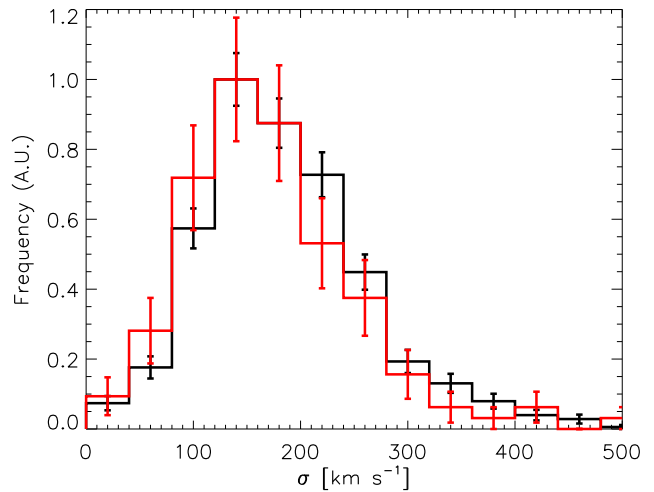}
\end{center}
\caption{Velocity dispersion distribution of the BOSS point source E+A (black) and complementary parent (red) samples. The error bars are Poisson errors.}
\label{fig:vdisp}
\end{figure}

To further investigate the link between compact quiescent galaxies and compact E+A galaxies, we examine the distribution of the central stellar velocity dispersion. We compare the velocity dispersion distribution of compact E+A galaxies {in the point source parent sample with the complementary point source parent sample, i.e. the point source parent sample with E+A galaxies removed.} Figure \ref{fig:vdisp} shows the velocity dispersion distribution of the E+A (black) and complementary parent (red) samples. 

We test whether the two distributions are drawn from the same underlying distribution using the Kolmogorov-Smirnov (KS) and Anderson-Darling (AD) tests. We limit the comparison to objects with velocity dispersions $70<\sigma<500$ km s$^{-1}$. The lower limit is set by the spectrograph resolution \citep{Thomas2013} and the upper limit rejects outliers. The KS and AD probabilities that the two samples are drawn from the same distribution are 21 and 19\%, respectively. Thus, the hypothesis that the data are drawn from the same underlying distribution can not be rejected. 

The velocity dispersion is a directly measured fundamental physical property of galaxies. {In the absence of merging, accretion or other dynamical interactions, the velocity dispersion of a quiescent galaxy should remain nearly constant as a function of time.} Thus, the similarity in the central velocity dispersion distributions of compact E+A galaxies and parent sample of compact quiescent galaxies supports the hypothesis that compact E+A galaxies are a progenitor of compact quiescent galaxies at intermediate redshifts.

\subsection{Number Density}

\begin{figure}
\begin{center}
\includegraphics[width=\columnwidth]{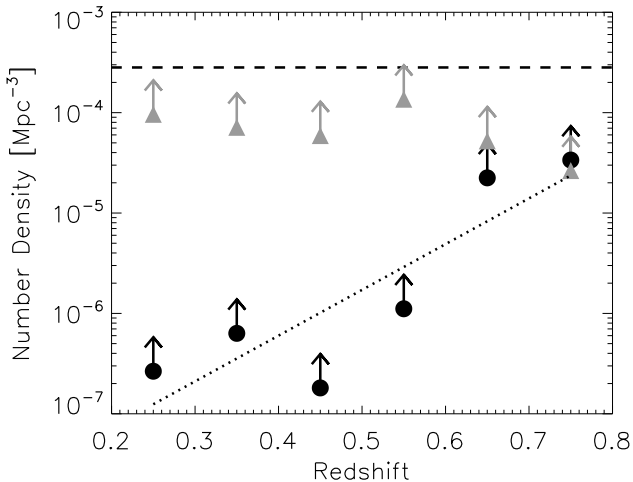}
\end{center}
\caption{Number density of compact E+A in the BOSS point source sample is shown in black points. The complementary sample of compact quiescent galaxies in the BOSS point source sample is shown by gray triangles. The data are corrected for spectroscopic incompleteness. They are lower limits to the true number density. {\citet{Damjanov2015a} measure the number density of compact galaxies (identified using the B13 classification) from publicly available data in the COSMOS field. They find that  the number density of compact galaxies is $10^{(3.55 \pm 0.26)}$ Mpc$^{-3}$; within the uncertainties, the number density is independent of redshift for galaxies at $0.2<z<0.8$. The dashed line shows the \citet{Damjanov2015a} number density estimate.} {The dotted line is a linear fit (Equation \ref{eq:fit_density}) to the lower limit of the number density of E+A galaxies.}}
\label{fig:density}
\end{figure}

\begin{figure*}
\begin{center}
\includegraphics[width=1.5\columnwidth]{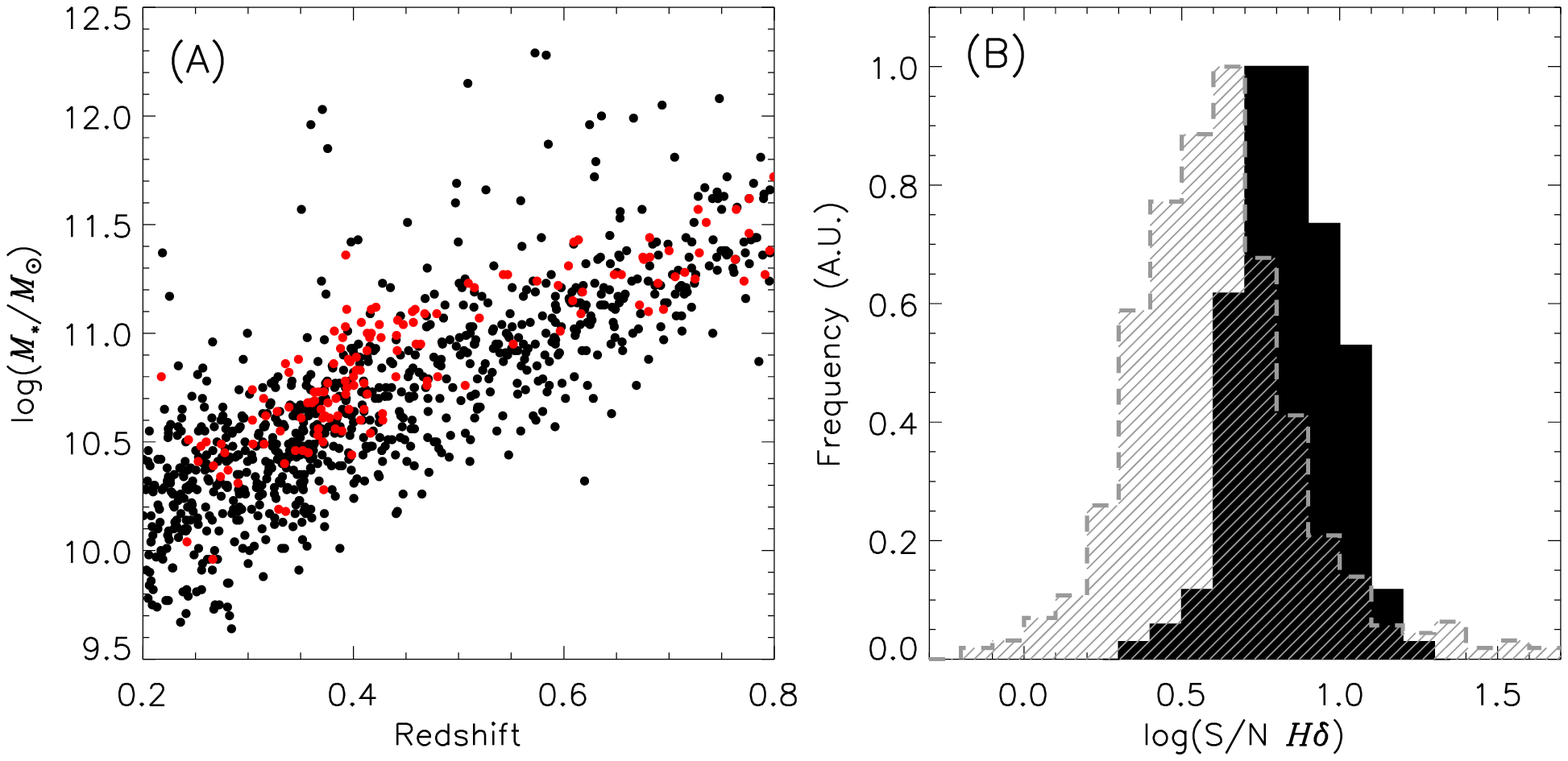}
\end{center}
\caption{(A) Stellar mass of the BOSS point source sample as a function of redshift. The black points show the parent sample and the red points are E+A galaxies. (B) The empirically determined S/N of the continuum measured in two windows adjacent to the H$\delta$ absorption line. The solid black and gray, hashed histograms are the S/N distributions for the BOSS point source and E+A and complementary parent samples, respectively.}
\label{fig:sn}
\end{figure*}

We next estimate the number density of E+A galaxies following \citet{Damjanov2014}. The BOSS data are sparsely sampled based on a hierarchy of selection criteria \citep{Ross2012}. Thus, we derive empirical corrections to account for spectroscopic incompleteness. We compare the total number of point sources in SDSS imaging with the number of objects spectroscopically targeted by BOSS as a function of magnitude and color. We derive the correction based on a 1000 square degree\footnote{The correction is derived from a fraction of the sample for reasons of computational efficiency.} subset of the full $\sim10,000$ square degree BOSS survey. We download the PSF magnitudes of all point sources in the SDSS imaging along with the flag indicating whether the object is targeted for spectroscopy. 

We derive the spectroscopic completeness by determining the fraction of galaxies targeted for spectroscopy relative to the total number of point sources as a function of $r$-band magnitude, $r-i$ and $i-z$ color. Following \citet{Damjanov2014}, we derive a completeness correction for each galaxy by taking a region of $\Delta r = 0.2$, $\Delta (r - i) = 0.4$ and $\Delta (i - z) = 0.5$ centered about the observed $r$-band magnitude, $r-i$ and $i-z$ colors, respectively. The region widths we adopt roughly corresponds to the typical $\pm 2\sigma$ confidence interval for the parent sample \citep{Damjanov2014}. In this region of magnitude and color space, we calculate the ratio of the number of spectroscopic targets to the total number of point sources. We weight the number count for each galaxy by this ratio. The median weight applied to the sample is $21.75$. A few objects inhabit a part of the color space where we can not derive a correction because of the scarcity of sources. For these objects, we apply no correction for spectroscopic incompleteness. The correction is insensitive to the width of the color and magnitude region or to the combination of colors used to derive the correction.

We estimate the number density in bins of redshift with $\Delta z = 0.1$. We take the total weighted number of galaxies over the cosmological comoving volume calculated from the fraction of the sky observed by BOSS and the redshift limits of each bin. In Figure \ref{fig:density} we plot the number density of BOSS galaxies as a function of redshift. {The black and gray points are the lower limits for the number density derived from the point source E+A sample and from the complementary (excluding E+A galaxies) point source parent sample, respectively.} The number density of unresolved sources remains relatively constant as a function of redshift whereas the the number density of E+A galaxies declines at late times. At $z\sim0.6 - 0.8$ the number densities of the two populations are comparable. {Because the number density estimates are lower limits, we are unable to draw any firm quantitative conclusions regarding the formation of compact quiescent galaxies at $z<0.8$. However, the analysis clearly demonstrates that some compact galaxies may first appear as quiescent at $z<0.8$, i.e. new compact quiescent galaxies form at intermediate redshifts.}

The lower limit to the number density of E+A galaxies {derived from the point source sample} as a function of redshift is
\begin{equation}
N_{E+A} (z)  = 10^{-5.8 \pm 0.2} (z - 0.5)^{4.6 \pm 1.3} \,\, \mathrm{ [Mpc ^{-3}]}.
\label{eq:fit_density}
\end{equation}
{We fit a linear model to the data in log-log space using \emph{linfit.pro} in IDL. We do not have an estimate of the observational uncertainty in the number densities because they are lower limits. The errors in the fit are determined from the deviation of the {lower limit number density estimates (black points in Figure \ref{fig:density}) from the linear model we fit. This procedure assumes that the linear model is the correct model of the data \citep{Press2002}.} The declining number density is in qualitative agreement with the declining fraction of E+A galaxies found in clusters as a function of time \citep[e.g.,][]{Dressler1983, Fabricant1991, Poggianti1999, Tran2003}.

\citet{Damjanov2014} show that the unresolved galaxies in BOSS provide a lower limit to the number density of compact galaxies. The dashed line in Figure \ref{fig:density} shows the number density of compact galaxies at intermediate redshifts based on HST imaging and highly complete spectroscopy \citep{Damjanov2015a}. The comparison suggests that the correction for spectroscopic incompleteness yields a lower limit that is typically a factor $\sim3-5$ below the measured value.

The estimates in Figure \ref{fig:density} are lower limits to the number density. Figure \ref{fig:sn}A shows the stellar mass as a function of redshift for the BOSS unresolved sample. The lowest stellar masses probed increases with redshift because the BOSS survey is magnitude limited. A similar declining trend is also present for the largest mass as a function of redshift. Size and stellar mass are correlated and the stellar mass where galaxies are unresolved in SDSS declines at lower redshift. Thus, the most massive compact galaxies are resolved at lower redshifts. These two effects contribute to incompleteness in the sample.

Figure \ref{fig:sn}A shows that E+A galaxies appear to follow similar trends to the parent BOSS sample but the stellar mass distribution shifts towards more massive galaxies. Figure \ref{fig:sn}B shows an empirical estimate of the S/N in the two continuum windows adjacent to the H$\delta$ absorption line. The distribution of E+A galaxies is offset to higher S/N. High S/N is required to identify E+A galaxies, therefore the E+A galaxies tend to be brighter, more massive objects than the parent BOSS sample.

\section{The Contribution of E+A galaxies to the Abundance of Compact Galaxies}

The properties of E+A galaxies in our sample are consistent with their being a progenitor of compact quiescent galaxies at intermediate redshifts. Here we ask whether the rate of compact E+A galaxy appearance is large enough to contribute significantly to the compact quiescent galaxy population. We derive a prediction for the rate of compact quiescent galaxy formation as a function of redshift assuming passive evolution. Specifically, we assume that: (1) compact E+A galaxies are the progenitors of compact quiescent galaxies which undergo a starburst, cease star formation and move onto the red sequence and (2) after the cessation of star-formation, compact E+A galaxies do not undergo any subsequent size growth.

The volume averaged rate of compact E+A galaxy formation depends on the volume number density of compact E+A galaxies and the timescale when the spectral properties that define E+A galaxies are visible. We calculate the lower limit to the number density of compact E+A galaxies as a function of time in Section 3.4. The E+A visibility timescale is limited by the evolving [OII]$\lambda3727$ and H$\delta$ EWs in post-starburst galaxies. Thus, the visibility timescale for an E+A galaxy depends on the timescale when a post-starburst galaxy meets the selection criteria; [OII] EW $< 2.5$ $\mathrm{\AA}$ and H$\delta$ EW $< -5 \mathrm{\AA}$. 

\citet{Falkenberg2009} model the spectral properties of E+A galaxy formation and evolution assuming various galaxy transformation scenarios and star formation histories. They adopt a less restrictive [OII] EW criteria for defining E+A galaxies ($<5$ $\mathrm{\AA}$). For their grid of models, the E+A phase is always $<0.7$ Gyr (see their Figure 9). Because of their weaker restriction on the [OII] EW and the fact that 0.7 Gyr is the longest timescale they find for the E+A phase, we adopt 0.5 Gyr as the typical visibility timescale. 

We combine the number density and visibility time of E+A galaxies to derive the volume averaged rate of compact quiescent galaxy production. The rate is
\begin{equation}
R(t) = \frac{N(t)}{\Delta T} \, \, \mathrm{[Mpc^{-3} \,\, Gyr^{-1}]}.
\label{eq:compact_rate}
\end{equation}
Here, $N(t)$ is the volume number density (shown in Figure \ref{fig:density}) and $\Delta T$ is the visibility timescale.

\begin{figure}
\begin{center}
\includegraphics[width=\columnwidth]{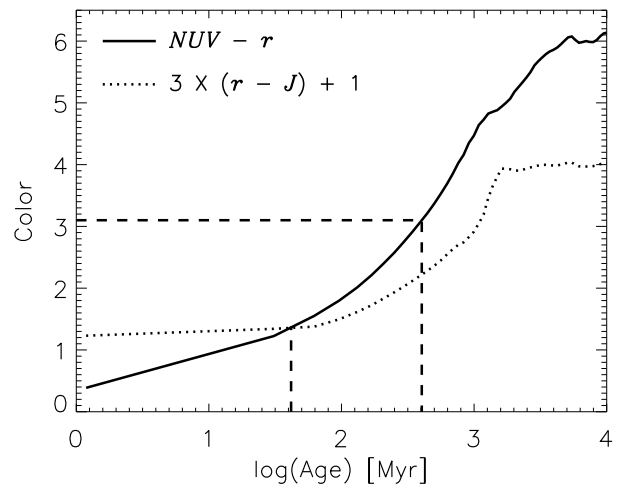}
\end{center}
\caption{Color evolution of a passively evolving galaxy as a function of time after the cessation of star formation based on the FSPS model. The $NUV-r$ and $3\times(r-J) + 1$ colors are plotted by the solid and dotted lines, respectively. The dashed lines at log(Age) $\sim1.6$ and $\sim2.6$ indicate when the passively evolving galaxy meets the $(NUV-r)>3\times(r-J) + 1$ and $(NUV-r)>3.1$ quiescent galaxy classification criteria of \citet{Ilbert2013}, respectively.}
\label{fig:red}
\end{figure}

A second issue for consideration when estimating the potential contribution of E+A galaxies to the compact quiescent galaxy population is the timescale of quiescence, i.e. the timescale over which galaxy properties evolve to the point where a galaxy is classified as quiescent. The quiescence timescale depends on the definition used to classify quiescent galaxies. We adopt the \citet{Ilbert2013} classification because it is based on the commonly used color-color diagram \citep[c.f.][]{Damjanov2015a}. \citet{Ilbert2013} classify galaxies as quiescent if $(NUV - r) > 3.1$ and $(NUV - r) > 3 \times (r-J)+1$. 

We quantify the timescale over which galaxy colors evolve by modeling a passively evolving galaxy with the Flexible Stellar Population Synthesis (FSPS) code \citep{Conroy2009a, Conroy2010}. We calculate $NUV -r$ and $r-J$ for a solar metallicity galaxy that undergoes a constant star formation rate which instantaneously shuts down after 500 Myr. Figure \ref{fig:red} shows the color evolution as a function of time after the shut down of star formation. $(NUV - r) > 3 \times (r-J)+1$ and $(NUV - r) > 3.1$ occur $\sim40$ and $\sim400$ Myr after the shut down of star-formation, respectively. Thus, we conclude that based on the \citet{Ilbert2013} classification galaxies rapidly come on to the red sequence after the shutdown of star formation. Because the quiescence timescale is comparable with the E+A visibility timescale and both timescales are an order of magnitude smaller than the fraction of cosmic time between $0.2<z<0.8$, we can assume that E+A galaxies instantaneously move onto the red sequence.

The predicted lower limit contribution of compact quiescent galaxies formed at $z<0.8$ to the compact quiescent galaxy population at intermediate redshifts is
\begin{equation}
N_c(t < t_{z=0.8}) \int_{t}^{t_{z=0.8}} R(t^\prime) dt^{\prime} \,\, \mathrm{[Mpc^{-3}]}.
\end{equation}
Here, $N_c(t)$ is the predicted number density of compact galaxy formation as a function of time $t$ for $z<0.8$, $t_{z=0.8}$ is the cosmic time at $z=0.8$ and $R(t^\prime)$ is the rate of compact galaxy production (Equation \ref{eq:compact_rate}). This calculation makes very simple, motivated assumptions; compact E+A galaxies instantaneously come onto to the red sequence (Figure \ref{fig:red}) and subsequently do not grow in size.

\begin{figure}
\begin{center}
\includegraphics[width=\columnwidth]{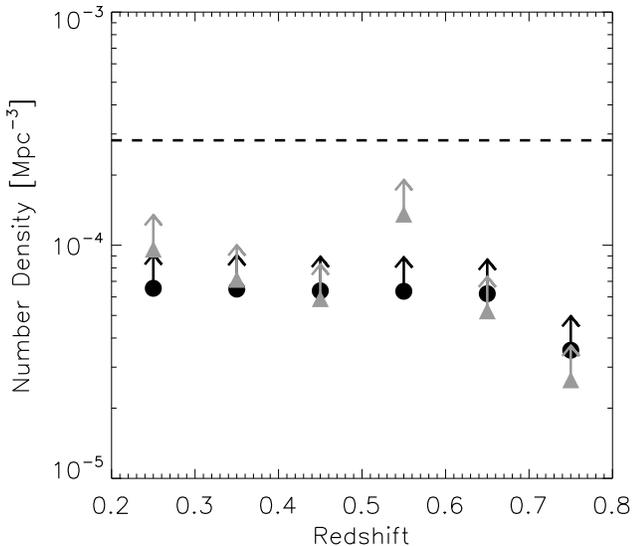}
\end{center}
\caption{Predicted number density of newly appearing compact galaxies is shown by the black points. The gray triangles are the number density of compact quiescent galaxies excluding compact E+A galaxies, same as in Figure \ref{fig:density}. The prediction is based on estimating a rate of compact galaxy production from the observed E+A galaxy number density as described in the text. The data have only been corrected for spectroscopic incompleteness and thus are lower limits to the true number density. The dashed line is the number density of compact galaxies measured in the COSMOS field \citep{Damjanov2015a}.}
\label{fig:rate}
\end{figure}

Figure \ref{fig:rate} shows the relative contribution of compact E+A galaxies to the older compact quiescent galaxy population at intermediate redshifts. The complementary compact quiescent galaxy sample shown in gray excludes young compact E+A galaxies. Thus, the complementary sample is mostly comprised of older galaxies (see Figure \ref{fig:age}). The combination of compact E+A galaxies and the complementary sample of compact quiescent galaxies makes up the total compact quiescent galaxy population seen in BOSS as point sources at $z<0.8$. The predicted lower limit contribution of compact galaxies first appearing as quiescent at $z<0.8$ is comparable to the lower limit number density of the complementary sample of compact quiescent galaxies. This suggests that {a large fraction} of the intermediate redshift compact quiescent galaxy population may first appear at intermediate redshifts. {Our analysis is based on BOSS which sparsely samples the galaxy population and uses a hierarchy of selection criteria; a complete spectroscopic survey with a simple selection function is necessary to quantify the formation rate of new compact galaxies at intermediate redshifts and to assess their relative contribution to the preexisting compact quiescent galaxy population. However, our analysis clearly demonstrates that} the intermediate redshift compact quiescent galaxy population is a mix of galaxies that have only recently shut down star formation and older galaxies which ceased forming stars at high redshift.

\section{Discussion}

Several studies based on the comparison of high redshift and local quiescent galaxies suggest that compact quiescent galaxies form at high redshifts and subsequently grow in an inside-out manner \citep[to cite a few]{Trujillo2007, VanDokkum2010a, Newman2012, Carollo2013, Toft2014}. The two observational cornerstones of this paradigm are the increase in the average size of compact galaxies at late times \citep[e.g.,][]{Daddi2005, vanderwel2014} and the rapid decline in the number density of compact galaxies in the local universe \citep{Trujillo2009, Taylor2010, vanderwel2014}. While the average growth in size may be explained by the addition of larger galaxies to the quiescent galaxy population, i.e. progenitor bias \citep{vanDokkum2001, Carollo2013, Cassata2013}, the purported dearth of compact galaxies in the local universe requires that a large fraction of the compact galaxy population observed at high redshift grows as a function of time. Regardless of whether the average growth of the population is due to growth of individual quiescent galaxies or to progenitor bias or both \citep{Belli2015}, the production of compact galaxies solely at high redshift remains central to the paradigm.

The compact galaxy formation and evolution paradigm largely connects measurements of the number density of compact galaxies at high redshifts with local measurements based on the SDSS \citep{Trujillo2009, Taylor2010}. One explanation for this reported precipitous decline in the number density is that quiescent compact galaxies somehow grow at intermediate redshifts. However, several studies examining dense regions in the local universe \citep{Valentinuzzi2010, Poggianti2013} and quiescent galaxies at intermediate redshifts \citep{Carollo2013, Damjanov2014, Damjanov2015a, Gargiulo2016} challenge these initial measurements of a rapid decline in the abundance of compact galaxy population at late times. These studies find that the number density of compact galaxies at $z<1$ is nearly constant. 

Motivated in part by the apparent rapid decline in the compact galaxy number density at late times, several possible growth mechanisms were proposed. Minor merger driven growth is a favored viable mechanism supported by observational evidence \citep{Newman2012}. However, the viability of this mechanism remains controversial \citep[e.g.,][]{Nipoti2009a, Nipoti2009b}. Thus, although the initial focus was explaining the inferred size growth of the quiescent galaxy population, the focus has shifted. The issue now is whether the constant number density of compact galaxies at $z<0.8$ can be reconciled with the growth and attrition of quiescent galaxies, whatever the mechanism.

We demonstrate that compact quiescent galaxies continuously form and that a significant fraction of the compact quiescent galaxy population at $z<0.8$ may first appear at these intermediate redshifts. Although this scenario does present a challenge to the dominant view that compact galaxies form exclusively at high redshifts, it helps to resolve the tension between growth of individual quiescent galaxies and the constant number density of compact galaxies at $z<1$. Moreover, our results explain the young stellar populations {($\lesssim2$ Gyr)} of some compact galaxies in the local and intermediate redshift universe \citep{Trujillo2009, Damjanov2014}. A constant number density of compact quiescent galaxies at $z<1$ requires that the rate of compact quiescent galaxy formation equals the rate of compact quiescent galaxy attrition due to the growth of individual compact galaxies. A better understanding of the physical processes responsible for the formation and growth of compact quiescent galaxies is necessary to explain this balance.

Two classes of objects$-$compact E+A and compact quiescent galaxies$-$may be linked by {a common formation scenario. The formation mechanism for these two types of objects remains uncertain but the major merger of two gas-rich galaxies offers one well studied possibility.} Simulations suggest that a high degree of dissipation is required to form compact remnants \citep[e.g.,][]{Cox2006, Khochfar2006} and large quantities of gas are needed to fuel a starburst. A gas-rich merger provides the physical conditions for triggering a starburst and for producing a compact remnant. {Thus, the E+A phase may be a common, short-lived evolutionary phase in compact galaxy formation and evolution. Depending on how compact quiescent galaxy samples are selected (e.g., photometric versus spectroscopic selection criteria), E+A galaxies may be included within compact quiescent galaxy samples or they may be progenitors of compact quiescent galaxies. In the latter case, passive evolution alone would be sufficient for them to be eventually classified as quiescent.} 

The models of \citet{Falkenberg2009} suggest that the post-starburst phase in a gas-rich merger scenario would be characterized by an E+A-type spectrum persisting for about $\sim500$ Myr. The compact E+A galaxies imaged with HST all have disturbed morphologies (see Figure \ref{fig:hst}), a result consistent with a merger scenario for E+A galaxies \citep[also see][]{Zabludoff1996, Yang2004, Goto2005}. Within this context, {some of} the compact star-forming galaxies examined by \citet{Diamond-Stanic2012} and \citet{Sell2014} may be the precursors of the compact E+A galaxies we study here. 

Compact quiescent galaxies may not form by a single mechanism. Gas-rich mergers may not be a necessary \emph{and} sufficient condition for the formation of compact galaxies and not all E+A galaxies are necessarily compact. Based on their physical properties, we simply show that compact E+A galaxies exist and they are naturally linked to the massive compact quiescent galaxies at $0.2<z<0.8$. Furthermore, the number density of compact E+A galaxies suggests that their passive evolution may contribute significantly to the compact quiescent galaxy population. Compact quiescent galaxies at $z<0.8$ have a wide range of stellar populations ages  \citep[see Figure \ref{fig:age};][]{Trujillo2009, Poggianti2013, Damjanov2014}. Thus, to maintain constant number density of compact quiescent galaxies at intermediate redshifts, there must be attrition and replenishment within the population.

\section{Conclusions and Future Prospects}

We identify 438 compact E+A galaxies in the SDSS and BOSS and make these data publicly available. This is the first large catalog of compact E+A galaxies. The mere existence of \emph{compact} E+A galaxies at $z<0.8$ means that some fraction of the compact quiescent galaxy population first appears at intermediate redshifts. We examine the sizes, stellar population ages and central velocity dispersions of these galaxies and demonstrate that compact E+A galaxies are a likely progenitor of compact quiescent galaxies at $0.2<z<0.8$. 

We derive a lower limit to the number density of compact E+A galaxies. We then construct a simple model assuming that compact E+A galaxies passively evolve. The model suggests that a substantial fraction of the $z<0.8$ massive compact quiescent galaxy population may first appear at these intermediate redshifts. These results link two classes of objects$-$compact E+A and compact quiescent galaxies$-$by a common evolutionary sequence.

The appearance of new compact quiescent galaxies at intermediate redshifts provides important constraints on theoretical models of galaxy formation and evolution. Major gas-rich mergers are a possible evolutionary mechanism linking compact E+As and compact quiescent galaxies. This scenario can be examined in more detail by combining state-of-the-art hydrodynamical models \citep[e.g.,][]{Cox2006} with simple stellar population synthesis techniques \citep[e.g.,][]{Falkenberg2009}. The constraints and formation scenarios for compact quiescent galaxies may also be investigated with large-scale cosmological simulations \citep[e.g.,][]{Wellons2015a}.

Quantifying the absolute number density of compact E+A galaxies is critical for understanding their role in the evolution of compact quiescent galaxies. Future measurements based on relatively complete, high quality spectroscopy in combination with high resolution HST imaging will advance towards this goal. These samples may be used to robustly quantify the distribution of stellar population ages of compact quiescent galaxies as a function of redshift. This approach measures the fractional contribution of newly appearing compact quiescent galaxies directly. This rate can then be used to evaluate the impact of size growth on the apparent abundance of compact quiescent galaxies at intermediate redshifts.

\acknowledgements

We thank the anonymous reviewer for comments that improved the clarity of the manuscript. H.J.Z. gratefully acknowledges the generous support of the Clay Postdoctoral Fellowship. N.B.H. was sponsored by the Smithsonian Astrophysical Observatory Latino Initiative Program. This program received Federal support from the Latino Initiatives Pool, administered by the Smithsonian Latino Center. I.D. is supported by the Harvard College Observatory Menzel Fellowship and the Natural Sciences and Engineering Research Council of Canada Postdoctoral Fellowship (NSERC PDF-421224-2012). M.J.G. is supported by the Smithsonian Institution. I.C. is supported by the Telescope Data Center, Smithsonian Astrophysical Observatory. His research related to compact galaxies is supported by the Russian Science Foundation project 14-22-0041; the RCSED catalog used in our study is supported by the grant MD-7355.2015.2 and Russian Foundation for Basic Research projects 15-52-15050 and 15-32-21062. 

We thank Charlie Conroy for assistance in implementing the FSPS modeling code. This research has made use of NASA's Astrophysics Data System Bibliographic Services and was partly based on observations made with the NASA/ESA Hubble Space Telescope and the Hubble Legacy Archive.

Funding for SDSS-III has been provided by the Alfred P. Sloan Foundation, the Participating Institutions, the National Science Foundation, and the U.S. Department of Energy Office of Science. The SDSS-III web site is http://www.sdss3.org/.

SDSS-III is managed by the Astrophysical Research Consortium for the Participating Institutions of the SDSS-III Collaboration including the University of Arizona, the Brazilian Participation Group, Brookhaven National Laboratory, University of Cambridge, Carnegie Mellon University, University of Florida, the French Participation Group, the German Participation Group, Harvard University, the Instituto de Astrofisica de Canarias, the Michigan State/Notre Dame/JINA Participation Group, Johns Hopkins University, Lawrence Berkeley National Laboratory, Max Planck Institute for Astrophysics, Max Planck Institute for Extraterrestrial Physics, New Mexico State University, New York University, Ohio State University, Pennsylvania State University, University of Portsmouth, Princeton University, the Spanish Participation Group, University of Tokyo, University of Utah, Vanderbilt University, University of Virginia, University of Washington, and Yale University.

\bibliographystyle{aasjournal}
\bibliography{/Users/jabran/Documents/latex/metallicity}

\begin{deluxetable*}{ccccccc}
\tablecaption{Properties of E+A Galaxies}
\tablehead{\colhead{RA} &\colhead{Dec} &\colhead{Redshift} & \colhead{$D_n4000$} &\colhead{log$(M_\ast/M_\odot)$ } & \colhead{$\sigma$ [km s$^{-1}$]} & \colhead{EW H$\delta$ [\AA]}}
\startdata
153.83205 & 1.06159 & 0.2158 & 1.29 & 11.25 & 180 $\pm$ 11 & -7.04 $\pm$  0.29 \\
\enddata
\label{tab:prop}
\tablecomments{The full catalog will be made available upon acceptance of the manuscript.}
\end{deluxetable*}

 \end{document}